\documentclass[sigconf]{acmart}
 
\usepackage{algorithm}
\usepackage{algorithmic}
\usepackage{booktabs}
\usepackage{threeparttable}
\usepackage{multirow}
\usepackage{subfigure}
\usepackage{graphicx}

\usepackage{booktabs} 
\usepackage{tabu}
\usepackage{multirow}
\usepackage{mathtools}
\usepackage{bm}

\usepackage{makecell}

\usepackage{xcolor}
\usepackage{soul}
\usepackage[utf8]{inputenc}
\usepackage{epsfig}
\usepackage{tabu}
\usepackage{multirow}
\usepackage{hhline}
\usepackage{setspace}
\usepackage{changepage}
\usepackage{amsmath}
\usepackage{comment}
\usepackage{multirow}
\usepackage{enumitem}

\usepackage{balance}
\usepackage{color}
\usepackage{amsmath}

\AtBeginDocument{%
  \providecommand\BibTeX{{%
    \normalfont B\kern-0.5em{\scshape i\kern-0.25em b}\kern-0.8em\TeX}}}

\setcopyright{acmcopyright}
\copyrightyear{2020}
\acmYear{2020}
\acmDOI{10.1145/1122445.1122456}

\acmConference[KDD '20] {The 26th ACM SIGKDD Conference on Knowledge Discovery and Data Mining}{August 22--27, 2020}{San Diego, CA, USA}
\acmBooktitle{The 26th ACM SIGKDD Conference on Knowledge Discovery and Data Mining (KDD'20), August 22--27, 2020, San Diego, CA, USA}
\acmPrice{15.00}
\acmDOI{10.1145/XXXXXX.XXXXXX}
\acmISBN{978-1-4503-6201-6/19/08}

\begin{document}

\title{CTR is not Enough: a Novel Reinforcement Learning based Ranking Approach for Optimizing Session Clicks}

\author{Shaowei Liu$^{\dagger}$, Yangjun Liu$^{\dagger}$}
\authornote{$~^{\dagger}$All the authors contributed equally to this research.}



\email{nphard6868@gmail.com, yankee_liu@outlook.com}

\renewcommand{\shortauthors}{S. Liu et al.}

\begin{abstract}
 Ranking is a crucial module using in the recommender system. In particular, the ranking module using in our online recommendation scenario is to provide an ordered list of items to users, to maximize the click number throughout the recommendation session for each user. However, we found that the traditional ranking method for optimizing Click-Through rate(CTR) cannot address our ranking scenario well, since it completely ignores user leaving, and CTR is the optimization goal for the one-step recommendation. To effectively undertake the purpose of our ranking module, we propose a long-term optimization goal, named as CTE (Click-Through quantity expectation), for explicitly taking the behavior of user leaving into account. Based on CTE, we propose an effective model trained by reinforcement learning. Moreover, we build a simulation environment from offline log data for estimating PBR and CTR. 
  We conduct extensive experiments on offline datasets and an online scenario. Experimental results show that our method can boost performance effectively.
\end{abstract}




\maketitle
\section{Introduction}
\begin{figure}[htpb]
\centering
\subfigure[]{
\begin{minipage}[t]{0.3\linewidth}
\flushleft
\includegraphics[width=1in]{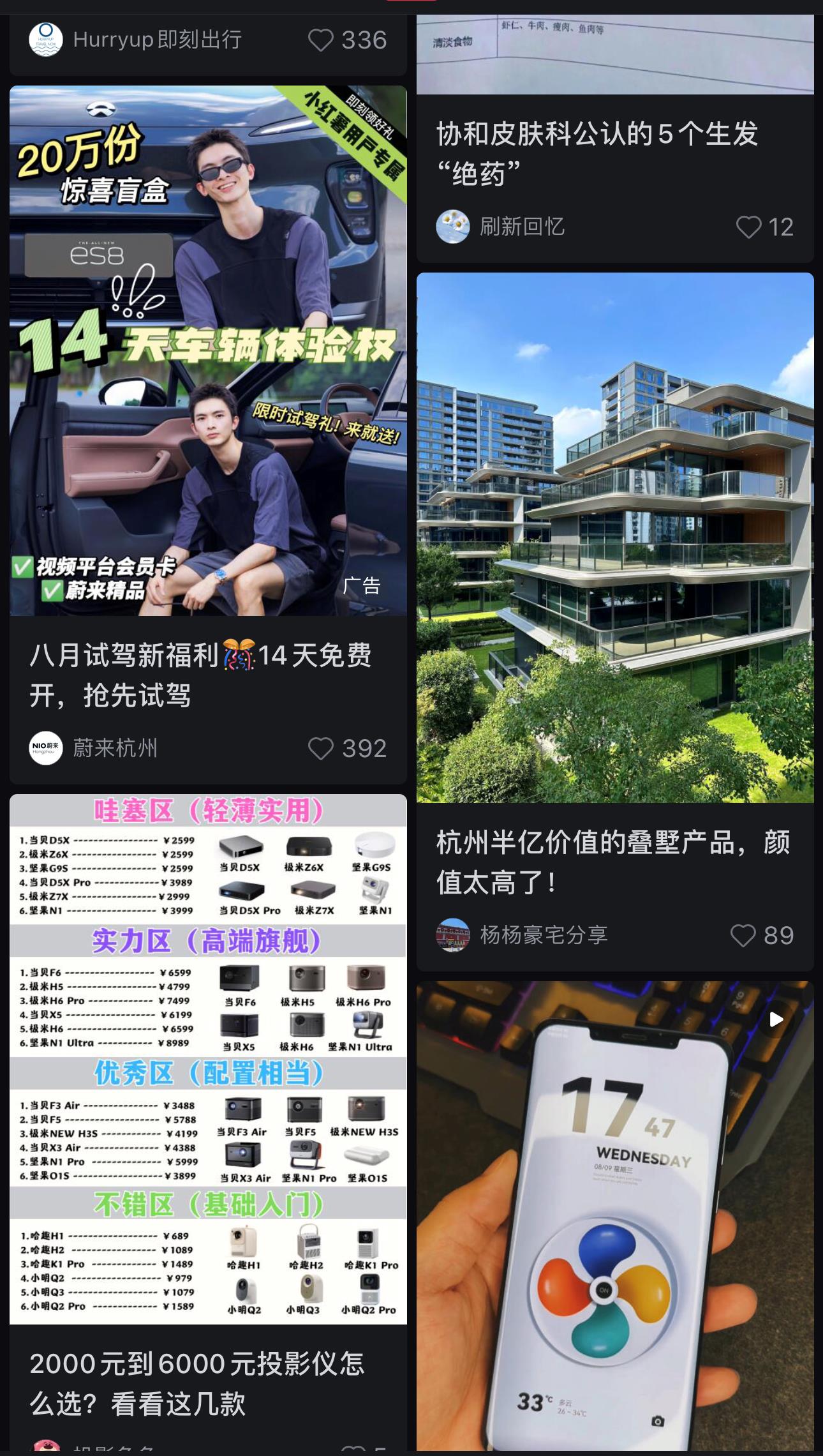}
\end{minipage}%
}
\caption{Illustraion of feed-streaming recommender system.}
\label{redbook}
\end{figure}

Feed-streaming recommender system is an emerging and important recommendation scenario, which can continuously deliver a never-ending feed-stream of items to the user. Due to the endless top-down presentation of items and easy operation by hand, feed-streaming recommender system is suitable for the user to browse contents or items on the mobile device constantly. Thus it is widely deployed on various mobile APPs such as Youtube, TikTok, and Taobao. Figurg (\ref{redbook}) shows one of feed-streaming recommendation scenarios, which can recommend multimedia contents (e.g., text, image, or video) generated by the other users. Specifically, the developed module exhibited in this paper is part of the entire  recommendation process, which is responsible for re-ranking a rough product collection obtained by the upstream system to make more accurate recommendations. Generally speaking, the business goal of those feed-streaming recommendation scenarios as well as our module is to attract users to continue staying in our feeds and consume as many reviews as possible by clicking them into detail page during users' browsing process. Therefore, our module should maximize the number of clicks of the user throughout her/his browsing session, rather than the click rate of a single product or a one-step recommendation.
 
During developing our module, we initially adopted traditional ranking methods to maximize the Click-Through Rate (CTR). However, we found that the user's browsing depth surprisingly declined while the CTR was improved. This finally leads to the decrease in click number. We assume that only focusing on CTR cannot guarantee that users stay in the system for a long time, and user total clicks cannot be maximized as users may leave the system prematurely. Thus, the user leaving should be explicitly taken into account for a recommender system with the purpose of maximizing the total amount of clicks. 


Specifically, we refer to the user's behavior of leaving a recommendation session as the user bounce. The probability of the user bounce at a certain position is denoted as Position Bounce Rate (PBR). By introducing the PBR, we can mathematically establish the connection between the user's leaving behavior and the user's browsing depth within a recommendation session. 
Further, a novel optimization goal, named as Click-Through Expectation (CTE), is derived from PBR and CTR to denote the expectation of accumulation of clicks throughout a recommendation sequence under the condition on the user's stay. Moreover, CTE is a long-term cumulative goal, which coincides with the purpose of reinforcement learning. Therefore, applying reinforcement learning should be an appropriate way in our recommendation scenario. 

In this paper, we propose a deep reinforcement learning (DRL) based method using a policy-gradient based learning algorithm, REINFORCE\citep{sutton2011reinforcement}, to optimize the long-term objective CTE. Recently, various DRL based recommendation models have been proposed\citep{zhao2018recommendations}\citep{ie2019slateq}\citep{DBLP:conf/kdd/ZouXDS0Y19}\citep{zhao2017deep}\citep{chen2019top}. Most existing works are based on the state transition triggered by the user's positive or negative feedback on the recommended item, e.g., click or rating. However, user's feedback cannot be acquired timely in our online environment because of the necessary process between upload data from user's device to online infrastructure. As a result, the feedback-based state transition is unrealistic for our scenario. Therefore, we propose a new MDP (Markov Decision-making Process) formulation based on the state transition driven by the user bounce. Distinguishing from the feedback, we can listen to the user bounce easily as it only happens at the end of the session, so that the transited new state can be quickly obtained to satisfy the time limit in our environment.  

Moreover, to address the high variance problem in REINFORCE, we introduce some reward baseline methods and involve them into REINFORCE to reduce the variance during training\citep{kool2019buy}. Besides, directly training the recommendation agent via the online environment is intractable and may damage users' experience. Hence, similar to the previous work, we first build a simulation environment from the off-line log data for estimating PBR and CTR. Then our recommendation agent is trained via the estimated CTR and PBR provided by the simulation environment.    

The main contributions of this work are summarized as follows:
(1) We found that only optimizing CTR cannot guarantee the improvement on the total number of clicks of a user. Based on the observation, we construct a novel training goal CTE to take the leaving behavior of users into account, and a new recommendation MDP formulation without the requirement of the user feedback. (2) We propose a DRL based model to optimize CTE in our recommendation scenario. To train the model more effectively and conveniently, we build a simulation environment from off-line log data. (3) We conduct numerous experiments on the off-line dataset and the online A/B test. The result shows that our method can significantly improve the user's click number, and the online experiments also demonstrate that our model is beneficial for the diversity of recommendation.

\section{Related Work}
\label{sec:related}

Recently, deep reinforcement learning has been applied to RS for optimizing LTV(Long Term Value). \citep{zhao2017deep} uses a DQN with the consideration of users' negative feedback. The MDP formulation of most existing DRL based recommendation models is similar to \citep{zhao2017deep} relying on user's feedback. \citep{zhao2018deep} expand their previews work using Actor-Critic framework to solve page-wise recommendation. \citep{chen2019top} use a corrected off-policy method to solve large-scale recommendation problem. In particular, \cite{DBLP:conf/kdd/ZouXDS0Y19} is a DRL based model to optimize the long-term user engagement (dwell time and revisit), which is highly similar to our goal for maximizing total click number of users. However, despite with 
same motivation and both two models have an simulation network in train process, but the target are different. Moreover, in MDP formulation, \cite{DBLP:conf/kdd/ZouXDS0Y19} assume that a user can immediately give feedback to every recommendation and before generating next one(like top one problem), which is unrealistic for our online environment settings. In contrast with existing works, our method focus on optimizing a single target, CTE and it's doesn't need users' feedback on each recommendation item.

 


\section{Proposed Method}
\subsection{Statistic Analysis}
\begin{figure}[htpb]
    \centering
    \includegraphics[scale=0.22]{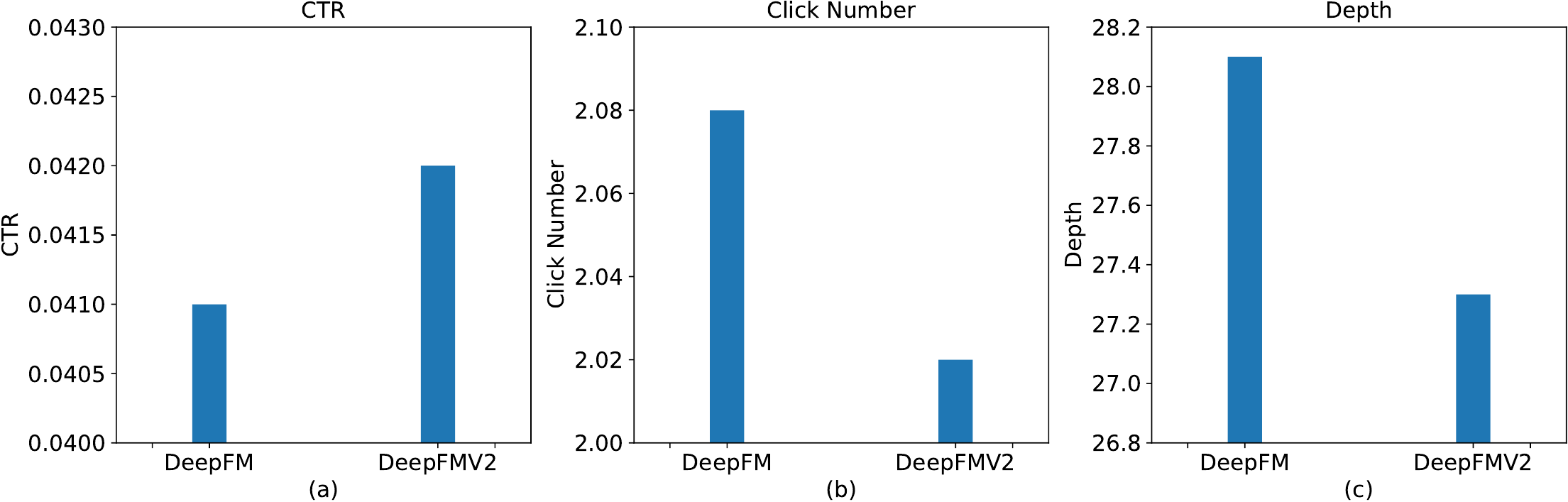}
    \caption{DeepFM \& DeepFMV2}
    \label{pic:motivation}
\end{figure}

We initially intended to improve the click number within a session by enhancing the adopted recommendation model's performance in terms of CTR directly. Specifically, the base recommendation model to be improved is DeepFM\citep{guo2017deepfm}, which is widely used in various recommendation scenarios. To further boost the performance of DeepFM in our scenario, we added more features to DeepFM that are more in line with our problem, e.g., users' thumbs-up, and we denote the improved DeepFM as DeepFMV2. To evaluate the performance of DeepFMV2, we applied DeepFMV2 to our online environment for one week A/B testing, and the results obtained by DeepFM and DeepFMV2 are exhibited in Figure (\ref{pic:motivation}). 

As shown in Figure (\ref{pic:motivation}(a)) and Figure (\ref{pic:motivation}(b)), the system's CTR can be improved by DeepFMV2 while the average click amount for different sessions decreases. According to Figure (\ref{pic:motivation}(c)), the worsen click number is correlated to the shortened browsing depth of users as early leaving of users. Higher CTR but shorter browsing depth is counter-intuitive. We assume this phenomenon is probably caused by the poor diversity of recommended products with high CTR or the recommendation being limited to the products that are highly relevant to the user's apparent interests for ensuring the maximization of CTR. Besides, the goal for optimizing the click number is long-term and accumulative, but CTR is immediate. As a result, only optimizing CTR can not guarantee the maximization of click numbers. Consequently, the behavior of user leaving should be explicitly considered, and we should use the optimization method for the long-term reward, e.g., reinforcement learning, instead of the conventional strategy for training recommender system with the immediate CTR goal.    

\subsection{CTE Optimization Goal}
Formally, we refer to the user's behavior of leaving a recommendation session as the user bounce. The overall goal of our system is to maximize Click-Through Expectation (CTE), which is correlated to CTR and PBR (Position Bounce Rate). Specifically, PBR is the probability of the user bounce at a certain position with a given item. The derivation of CTE is as follows. 

Let $\mathcal{I}$ denote the set of items to be recommendation. At each position $t$, the user sees the recommended item $i_t\in\mathcal{I}$. Let $C(t,i_t)$ be the user's click feedback on $i_t$, that is, $C(t,i_t)$ is 1 if user clicks the item and 0 otherwise. Without loss of generality, let $T$ be the maximum number of feeds in one session (as a longer session can be decomposed into several smaller ones). For any given ordered item list $(i_1, \dots, i_T)$, the total number of clicks in one session is given by $\sum_{t=1}^{T} C(t,i_t)$. Our goal is to maximize its expectation version $CTE(i_1, \dots, i_T)=E\sum_{t=1}^{T} C(t,i_t)$ over all possible ordered item list with length $T$. 

For each recommendation, directly choosing $T$ items from $\mathcal{I}$ with top CTR, however, may not maximize CTE in feeds recommendation as a user may leave at a early stage. To this end, we introduce the conditional CTR and PBR to explicitly model a user's click and bounce behavior. Specifically, let $RI = \{(1, i_{1}), (2, i_{2}), \cdots, (T, i_{T})\}$ be our recommendation sequence and $\mathcal{H}_{t}=\{(1,i_1),(2,i_2),\dots,(t,i_t) \}$ be the presumed browsing history for a user browse until position $t$, and $B(t, i_t) \in \{0,1\}$ represents the user's bounce feedback at position $t$. Note that $\mathcal{H}_t$ implies $B(k,i_k)=0$ for $k=1,\dots,t-1$, which means the user continues to browse at position $1,\dots, t-1$. Now the conditional CTR and PBR is given by
\begin{gather*}
CTR(u,t,i_t,\mathcal{H}_{t}) = P(C(t,i_t)=1|u,\mathcal{H}^u_t), \\
PBR(u,t,i_t,\mathcal{H}_t) = P(B(t,i_t)=1|u,\mathcal{H}^u_t)
\end{gather*}
We emphasize that CTR and PBR are both conditional probabilities, which not only relies on the fact that a user not leaving until position $t$, but also is context-aware by considering historical impressions. For notational simplicity, we will use $CTR_t$ and $PBR_t$ and omit other arguments in the following when there is no ambiguity. 

With the definition of CTR and PBR, we are ready to re-express CTE using CTR and PBR by Bayes' rule. Specifically, for any given ordered ranking list $(i_1, \dots, i_T)$, we have
\begin{equation}
   \begin{aligned}
    \label{CTE_CTR_PBR}
    &CTE(i_1, \dots, i_T) = E \sum_{t=1}^T C(t,i_t) = \sum_{t=1}^T P(C_{t}=1, \mathcal{H}_t) + P(C_{t}=1, \mathcal{H}_t^c) \\
    &= \sum_{t=1}^T P(C_{t}^{1}) P(\mathcal{H}_{}|\mathcal{H}_{t-1}) P(\mathcal{H}_{t-1}) = \sum_{t=1}^T CTR_t\cdot P(B_{t-1}^{0})P(\mathcal{H}_{t-1}) \\
    &= \sum_{t=1}^T CTR_t \cdot (1-PBR_{t_1}) \cdot P(\mathcal{H}_{t-1}) = \sum_{t=1}^T CTR_t \prod_{k=1}^{t-1}(1-PBR_k),
    \end{aligned}
\end{equation}

where $P(C_{t}^{1}) = P(C(t, i_{t}=1)|\mathcal{H}_{t})$, $P(B_{t}^{0}) = P(B(t, i_{t})=0|\mathcal{H}_{t})$ and the last equation follows by iteratively expanding $P(\mathcal{H}_t)$.

Even if armed with machine learning models to predict CTR and PBR, directly optimizing (\ref{CTE_CTR_PBR}) is infeasible in practice, as it's impossible to enumerate all possible ordered ranking list to compute its corresponding CTE. Therefore, we propose to reformulate this problem using a reinforcement learning framework, where the recommender system interacts with the user by feeding an item $i_t$ at position $t$, observes a new state (bounce or not), and receives the reward (click or not). The ultimate goal is to maximize the cumulative reward.

\subsection{MDP Formulation of Ranking Process}

Reinforcement learning is based on MDP\citep{sutton2011reinforcement}, thus we first formulate the MDP using in our recommendation scenario. 

A MDP is defined by $M = \langle S, A, P, R, \gamma \rangle$, where $S$ is the state space, $A$ is the action space, $P: S \times A \times S \rightarrow$ $\mathbb{R}$ is the state transition probability, $R: S \times A \rightarrow$ $\mathbb{R}$ is the reward function, and $\gamma \in [0,1]$ is the discount factor. Besides, a policy $\pi: S \times A \rightarrow [0, 1]$ is a distribution over actions given states, where $\pi(a|s)=P(A_t=a|S_t=s)$. In our scenario $\langle S, A, P, R \rangle$ are set as follows.

\textbf{State}. A state $s_t$ describes the process status, which includes the available items $\mathcal{I}$, user profile $u$, the presumed browsing history $\mathcal{H}_t$, and the user's bounce feedback $B(t,i_t)$. At the beginning, we have $s_0 = \{u, \mathcal{I}, \emptyset, 0\}$, since no item has been exposed and we assume the user will see the first item for sure. We presume the user continues to browse, his state at time step $t$ is given by $s_t = \{u, \mathcal{I}, \mathcal{H}_{t-1}, B(t,i_t) \}$. Note that when $B(t,i_t)=1$, the recommend session comes to a terminal state and no more actions could be taken.
 I 
	
\textbf{Action}. Let $\mathcal{I}_t=\{ i| (j,i)\in\mathcal{H}_t \text{~for some~} 1\leq j \leq t \}$ be the set of items already recommended until position $t$. At each state $s_t$ satisfying $B(t,i_t)=0$, we could choose an item $a_t \in \mathcal{I}\setminus \mathcal{I}_t$ to appear at position $t+1$.
	
\textbf{State Transition}. After we take an action $a_t$, the browsing history for a user is appended by $(t, a_t)$, and $s_{t+1}$ is determined by $PBR_{t+1}$ by its definition. That is, 
	\begin{align*}
	P(s_{t+1}=\{ u,\mathcal{H}_t\cup\{(t+1, a_t)\},B(t,i_t)=1 \}|s_t, a_t) &= PBR_{t+1}, \\
	P(s_{t+1}=\{ u,\mathcal{H}_t\cup\{(t+1, a_t)\},B(t,i_t)=0 \}|s_t, a_t) &= 1-PBR_{t+1}.
	\end{align*}
	
\textbf{Reward}. At each step $t$, the reward $R(s_t,a_t)$ is the user's click feedback on item $(t+1, a_t)$, that is, $C(t+1, a_t)$. Hence the return $G_t$, the cumulative discounted reward from time $t$, is given by 
	\begin{equation}
	G_t = \sum_{t=0}^{T-1} R(s_t, a_t) = \sum_{t=0}^{T-1} C(t+1, a_t).
	\label{equ:reward}
	\end{equation}

If we assume for simplicity that $\pi(a|s)$ is a deterministic policy that generates the ordered ranking list $(i_1,\dots,i_T)$, the state-value function $v_\pi(s_0)$ becomes
\begin{align*}
v_\pi(s_0) &= E_\pi(G_0|s_0) 
= \sum_{a_0\in \mathcal{I}} \pi(a_0|s_0) \left(R(s_0,a_0) + \gamma \sum_{s_1 \in S} P(s_1|s_0) v_\pi(s_1) \right) \\
&= CTR_1 + \gamma (1-PBR_1) v_\pi(s_1) = \sum_{t=1}^T CTR_t \prod_{k=1}^{t-1} \gamma^k (1-PBR_k),
\end{align*}
which is exactly CTE defined as in (\ref{CTE_CTR_PBR}) if $\gamma=1$. As CTR and PBR are both conditional probabilities depending on the browsing history, thus affected by the policy $\pi(a|s)$, our goal is to seek a policy that maximizes the initial state-value function.

To this end, we have constructed the MDP for our recommendation scenario. The remaining issue is how to get the policy function $\pi(a|s)$, and we employ neural networks to approximate the policy function with a simulation environment for training the neural network. The related details are presented in next section.

\section{Architecture \& Learning }
\label{sec:model}


\subsection{Policy Network}

The neural network for policy $\pi(a|s)$ is required to take state $s$ as input and output the action distribution $\mathcal{P}(I) \sim \pi_{\theta}(s)$, and then the recommended item or action is sampled from the $\mathcal{P}(I)$. As shown in previous section, state $s_t = \{u, \mathcal{I}, \mathcal{H}_{t}, B(t,i_t) \}$, where $u$ and $\mathcal{I}$ are the explicit inputs of our neural network, $\mathcal{H}$ is viewed as a implicit input, but $B(t, i_{t})$ isn't considered as a input as it is only used to determine whether the session is terminated. 

\begin{figure}[tb]
    \centering
    \includegraphics[scale=0.3]{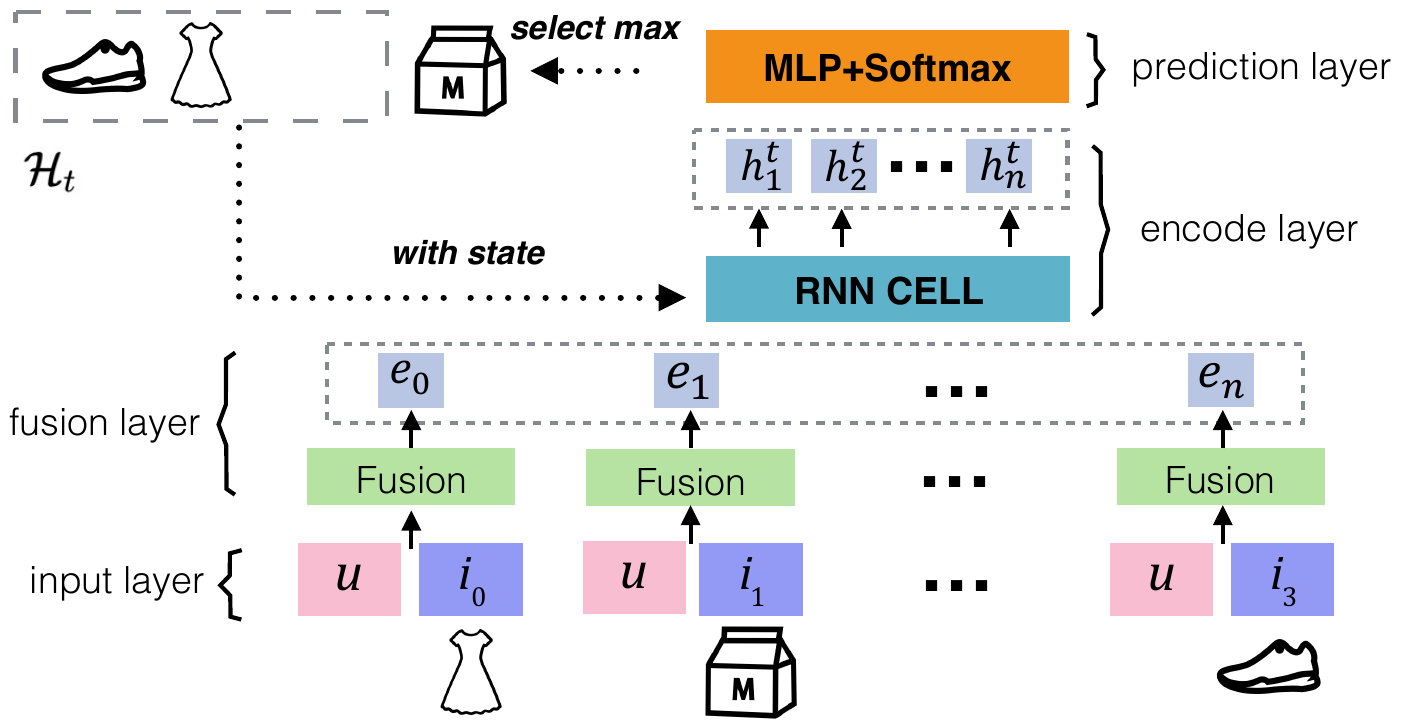}
    \caption{The architecture of rank policy}
    \label{fig:agent}
\end{figure}

The overall neural network architecture is shown in Figure~\ref{fig:agent}, which includes three major components: the fusion layer for the explicit input, the encoding layer for the implicit input and the prediction layer for obtaining the action distribution. The details of different layers are given as follows. 

\subsubsection{Fusion layer}
Suppose that item $i \in \mathcal{I}$ is in the candidate set $\mathcal{I}$, $N=| \mathcal{I}|$ is the size of $\mathcal{I}$,  $\hat{i}=\{x_1^i,x_2^i,...,x_n^i\}$ is a vector consisting of various side information features for item $i$, and $\hat{u}=\{x_1^u,x_2^u,...,x_n^u\}$ is a vector including the features of user $u$. 

We employ a FM (Factorization Machine) with a 3 layers MLP (Multi-Layer Perceptron) to generate a high-level representation $e_{i}$ for each pair of $\hat{u}$ and the item  $i \in \mathcal{I}$, as follows, 
\begin{equation}\label{fusion_layer}
\begin{split}
    e_{i} = MLP(FM(\hat{u},\hat{i})),\ \ \ 
    E  = \{e_{1}, e_{2}, ..., e_{i},..., e_{N} \}.    
\end{split}
\end{equation}
where $E$ is a matrix for $\mathcal{I}$ and $e_{i}$ is the column vector of $E$. To be more clear, we only use FM to get crossed features vector between different fields. Namely, the FM features is, $\{x_1^u x_1^i, x_1^u x_2^i, ..., x_1^u x_n^i, ..., x_n^u x_n^i \}.$

Note that $\mathcal{I}$ doesn't include all products in our application, but only contains a relatively small item set obtained by an upstream ranking process. Therefore, our recommendation agent actually plays a re-ranking role to further retrieve more valuable products from a rougher candidate set, and thus handling all items in $\mathcal{I}$ is tractable. In addition, $\mathcal{I}$ is consistent throughout the entire recommendation session for a user.  

\subsubsection{Encoding layer} 
We first utilize a RNN with GRU cell to further encode the representation $e_{i}$ in matrix $E$, where the length of the RNN is $N$. Specifically, the column vectors in the matrix $E$ are sequentially feed into the RNN. This procedure is denoted as follows,
\begin{equation}
\label{equ:gru}
\begin{split}
    h^t_{i} = RNN(e_{i}, h^{t-1}_{N}), \ \ \ 
    O^{t}_{H} = \{ h_{1}^t, h_{2}^t, ..., h_{N}^t \}  
\end{split}
\end{equation}
where $O^{t}_{H}$ is the matrix including all RNN hidden features for items in $\mathcal{I}$ and $h^{t-1}_{N}$ is the RNN hidden features of a chosen item in last step.

Moreover, $O^{t}_{H}$ implies the pattern of recommendation at time $t$ since the following item selection is based on $O^{t}_{H}$, and the selected item $i_{t}$ will be appended into  $\mathcal{H}_{t-1}$ to acquire updated history $\mathcal{H}_{t}$, i.e., $\mathcal{H}_{t} = \mathcal{H}_{t-1} + \{ i_{t} \}$. Namely, $\mathcal{H}_{t}$ is highly correlated to $O^{t}_{H}$. As a result, we use $O^{t}_{H}$ to denote $\mathcal{H}_{t}$ implicitly, instead of explicitly taking $\mathcal{H}_{t}$ as input. The advantage of this way is by directly reusing the RNN hidden state in last step, it can save much computing time since our online serving has a strict limitation on runtime.

We let the initial hidden features $h_{0}^{t-1}$ and $h_{0}^{t}$ are different to ensure nonidentical $O^{t-1}_{H}$ and $O^{t}_{H}$, so that the recommendation patterns at $t-1$ and $t$ are variable. Concretely, the initial hidden features $h_{0}^{t}$ is obtained as follows,
\begin{equation}
\label{equ:gru1}
\begin{split}
    h^t_{0} = h^{t-1}_{N},\ \ \ 
    h^1_{0} = \text{Zeros vector}.
\end{split}
\end{equation}
Besides, the assignment operation, i.e., $ h^t_{0} = h^{t-1}_{N}$, also can be viewed as the state transition from state $s_{t-1}$ to state $s_{t}$.  


\subsubsection{prediction layer} 
The final action distribution for the policy $\pi_{\theta}(a|s)$ at step $t$ is obtained from $O^{H}_{t}$ via softmax operation, i.e. $
    \pi_{\theta}(a | \mathbf{s})=\frac{\exp \left({W}_{s} \mathbf{h}_{i}^{t} \right)}{\sum_{\mathbf{h}_{i^{\prime}}^{t} \in \mathcal{H}} \exp \left(W_{s} \mathbf{h}_{i^{\prime}}^{t}\right)}$, 
where $B(t, i_{t})$ in state is not directly participate in inference process, it used for noticing the policy that user has left the system.

\subsection{Simulation Environment}
Directly training the recommendation agent in the online environment would increase the burden of the online infrastructure and damage the user experience. Therefore, we further build a simulation environment (SE) for estimating CTR and PBR, which is trained through offline logs. SE is similar to the policy networks, which also contains a fusion layer, a encoder layer and a prediction layer, as shown in Figure~\ref{fig:environment}. The details are given below.
\begin{figure}[tb]
    \centering
    \includegraphics[scale=0.3]{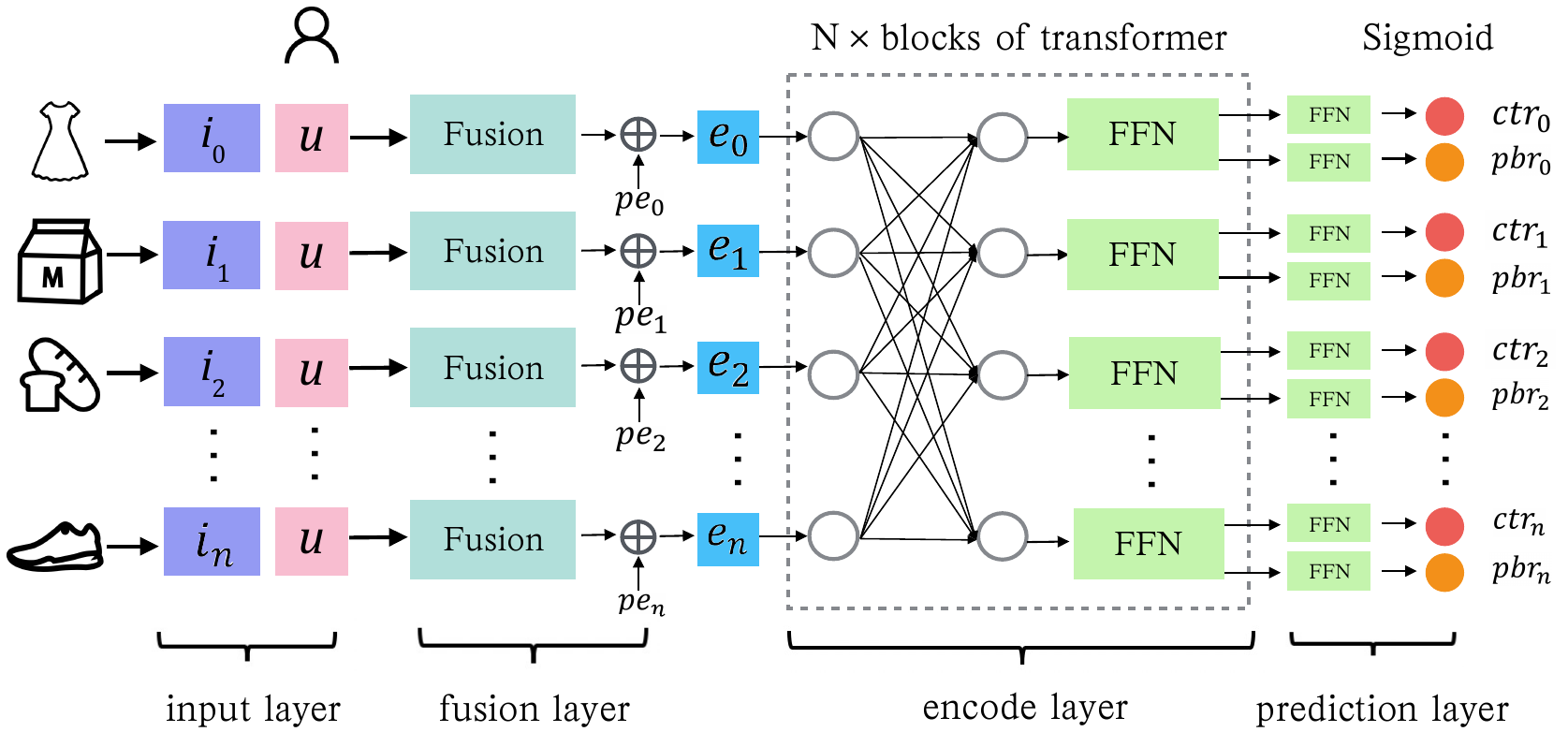}
    \caption{The architecture of simulation environment}
    \label{fig:environment}
\end{figure}

\subsubsection{Fusion layer}
The architecture of the fusion layer in SE is consistent with the one using in the policy networks, but the input is $H_{t}$ instead of $I$ for the policy networks. Specifically, the output matrix of the fusion layer is $\hat{E}=\{\hat{e}_1,\hat{e}_2,...,\hat{e}_t\}$, and its column vector $\hat{e}_j$ is obtained by $\hat{e}_j = MLP(FM(\hat{u},\hat{j}))$, where $\hat{j}$ is the feature vector of item $j \in H$. 

\subsubsection{Encoding layer}
We use a Transformer instead of the RNN employing in the policy networks as the encoder layer of SE since off-line training environment is able to afford more powerful and time-consuming model likes Transformer. 

The Transformer layer can be denoted as, 
$
    \hat{h}_{i} = Transformer(\hat{E}, P),  
$
where $P=\{pe_{1},pe_{2},...,pe_{m} \}$ is a trainable matrix including the position vector $pe_{i}$ for each item to eliminate the position bias.

\subsubsection{Prediction layer}
The final output of SE is the scalar $CTR_{t}$ and the scalar $PBR_{t}$, which can be generated from $\hat{h}_{i}$ via a MLP, i.e. $(CTR_{t}, PBR_{t}) = MLP(\hat{h}_{i}).$

The cost function of SE is,
\begin{equation}\label{environment_loss}
    \mathcal{L}_{SE} = -\sum_{i=1}^{n} y_{ci}log(P(y_{ci}|u, \mathcal{H}_{i}, \theta_{ci})) - \sum_{i=1}^{n} y_{bi}log(P(y_{bi}|u, \mathcal{H}_{i}, \theta_{bi}))
\end{equation}
where $y_{ci}$ is the CTR label indicating whether the user clicks the item at position $i$ of the sequence, $y_{bi}$ is the PBR label indicating whether the user leaves at position $i$ and $n$ is the number of items in the sequence. The simulation environment is trained with log labeled data and optimized by mini-batch SGD\footnote{We provide our model file at https://github.com/qqhard/RL4CTE}.

\subsection{Reinforce Policy Training}

We update our policy by interacting with the simulation environment.
The algorithm involves generating a complete episode and using the accumulated rewards to optimize the networks. The complete algorithm is shown in Algorithm~\ref{alg:Framwork}.

In this work, we focus on a policy-gradient-based approach REINFORCE. The gradients of the cumulative reward with baseline $G_{t}^{\prime}$ w.r.t. the policy parameters can be derived with logarithmic function in Equation~\ref{equ:loss_baseline},
\begin{gather}
\label{equ:loss_baseline}
    \nabla J(\theta)= G_{t}^\prime \nabla \log \pi\left(a_{t} | s_{t}, \theta\right).
\end{gather}

We use the baseline $G_{t}^{\prime}$ instead of the basic accumulative rewards $G_{t}^{j}$ for reducing the variance of gradients in REINFORCE. Specifically, we consider two kinds of baseline functions as follows.

\textbf{Whitening}. This method is to scale the returns of each step via the mean and standard deviation within the episodes, as shown in Equation~\ref{Whitening},
    \begin{equation}
    \label{Whitening}
    G_{t}^{\prime} = \frac{G_{t} - \hat{G}}{\sigma G}.\\
    \end{equation}

 \textbf{Sampled Baseline\ref{kool2019buy}}. We first perform $N$ trajectories containing $k$ steps  for each state, and then we use the average return as the baseline. Sampled Baseline is formally denoted as, 
    \begin{equation}
    \label{re_baseline}
        G_{t}^{j\prime} = G_{t}^{j} - \frac{1}{N-1}\sum_{b \neq j}G_{t}^{b},
    \end{equation}
    where $G_{t}^{b}$ is the cumulative reward at step $t$ in $b$-th trajectory.
    

\begin{algorithm}[htbp]
\caption{Re-ranking model with Simulation Environment}
\label{alg:Framwork}
\begin{algorithmic}[1] 
\REQUIRE ~~\\ 
  Offline train data $\mathcal{D}$; Rank policy $\pi_{\theta}$; 
  Rank size $k$; Replay Buffer $\mathcal{M}$;
  Simulation environment as $S_{\theta}$; \\
\ENSURE ~~\\ 
\STATE \textbf{\# train simulation environment (SE)}
\REPEAT
\STATE Sample random mini-batches of $(u, H_{t-1}, i_{t}, y_{ci}, y_{bi})$ from $\mathcal{D}$\\
\STATE Update $S_{\theta}$ via mini-batch SGD using equation \ref{environment_loss}
\UNTIL convergence
\STATE \textbf{\# train policy network}
\REPEAT
\STATE $\mathcal{M} \leftarrow \varnothing $
\STATE Sample mini-batches $\mathcal{B}$ of $(u, I, \varnothing)$ from $\mathcal{D}$
\FOR{$b$ in $B$}
\STATE $b=(u, I, \varnothing )$ $\in$ $\mathcal{B}$
\REPEAT
\FOR{t from 1 to $k$}
 \STATE Drawn an item $a_{t}$ from $\pi_{\theta}(s_{t})$, where $s_{t}$ = $[u, I, \mathcal{H}_{t-1}]$\\
 \STATE Update state $s_{t+1} = [u, I, \mathcal{H}_{t-1} \cap a_{t}]$ and get $CTR$, $PBR$ of $a_{t}$ from SE.
 \STATE    Store $(s_{t}, s_{t+1}, a_{t}, CTR_{t}, PBR_{t})$ in $\mathcal{M}$
\ENDFOR
\UNTIL $\mathcal{M}$ has $N$ trajectories
\STATE Calculate cumulative reward $w.r.t$ Equation \ref{equ:reward}, get the baseline $G_{t}^{\prime}$ $w.r.t$ Equation \ref{re_baseline} and store in $\mathcal{M}$
\ENDFOR
\STATE Sample a mini-batches from $\mathcal{M}$
\STATE Update the policy $\pi_{\theta}$ $w.r.t$\\ $\theta_{t+1}=\theta_{t}+\alpha G_{t}^{\prime} \nabla \log \pi\left(A_{t} | S_{t}, \theta\right)$ 
\UNTIL convergence
\end{algorithmic}
\end{algorithm}


\section{Experiments}
\label{sec:experiment}










\subsection{Dataset}
\label{sec:dataset}
We choose two representative datasets to evaluate our method.

\textbf{Online dataset.} This dataset collected from our own scenario. Samples for 30 days contain 160 million sessions which covers 10 million users and 5 million items. We log recommended item, user id, users' behavior,  time and so on to rebuild sessions off-line.

\textbf{Yahoo.} Yahoo is a public dataset for Yahoo! Learning to Rank Challenge, which includes 25,000 samples. Each sample contains a vector of 700 dimensions representing the doc-query pair with a rating score in (0,1,2,3,4). Unfortunately, neither users' browsing history nor bounce position is included in the dataset. So we construct a dataset in the pre-processes proposed in ~\cite{Joachims2018} according to Yahoo dataset. In the following, we introduce the pre-processes in more details.

The pre-processes are divided into two major stages. First, we generate the exposure sequences to simulate users' browsing and click behaviors. Second, we insert the bounce label into the exposure sequences. The procedure is shown as follows.

 \textbf{Exposure sequences.} We adopt a click data generation method like ~\cite{Joachims2018} to generate exposure sequences. More specifically, first train a ranking model MART with the raw data, note that an item is labeled as clicked if its rating score is above 2 which is consistent with the original data. Exposure sequences are obtained by sorting items according to ctr from MART.
\textbf{Bounce label:} We suppose that a user will leave a session when the user's interests are not satisfied for a period of time. Therefore, the measurement of the user's satisfaction at a given position should consider not only the present item but also the results of previous recommendations. To capture user's interest, we defined a modified MMR (Maximal Marginal Relevance) as follows, where $\lambda$ is 0.1:
    \begin{equation}
    \label{MMR}
    MMR_j {=} \lambda Rating_j + (1-\lambda) \min _{I_{j} \in S} \operatorname{Dis}\left(I_{i}, I_{j}\right)
    \end{equation}
  where $j$ indicates the j-th item in the exposure sequence, $\lambda$ is the weight of CTR and used to represent the influence of CTR, $Rating_{j}$ represents the predict CTR of the j-th item from MART, S is the set of exposed items, $I_{i}$ is the $i$-$th$ item in S, and $\operatorname{Dis}\left(I_{i}, I_{j}\right)$ is the Euclidean distance between item i and j where the item vector is extracted from MART. Based on the modified MMR, the cumulative satisfaction of a user is the average MMR over the exposure sequence formulated as follows, 
  \begin{equation}
 \label{DecayRate}
 {Decay Rate}_j {=} \frac{1}{j} \sum_{i=1}^{j} {MMR_i}
 \end{equation}
  Finally, the the bounce label ${BounceLable}_{j}$ of jth item can be obtained as follows. Here we set $threshold$ 0.8.
\begin{equation}
\label{bounceLable}
{BounceLable}_{j} = \left\{
             \begin{array}{lcl}
             1 & {Decay Rate}_j < threshold\\
             0 &\text{otherwise}
             \end{array}.
        \right.
\end{equation}

\subsection{Evaluation Metrics}
The number of clicks in a session and the browsing depth are the most importation targets in the feeds RS, where the number of clicks and the browsing depth indicate users' interests and stickiness, respectively. In both of the off-line and on-line experiments, the number of clicks and the browsing depth are used as the measure metrics, presented as follows.

\textbf{Average Clicks Per Session(AC).} The average cumulative number of clicks of sessions.
\begin{equation}
  AC=\frac{1}{N}(\sum_{i=1}^{N}\sum_{j=1}^{D}click_j)
\end{equation}
where $N$ is the number of sessions, $D$ is the depth of a session which is equal to the bouncing position, and $click_{j}$ is 1 if a user clicks the j-th item.

\textbf{Average Depth Per Session(AD).} The average browsing depth of sessions.
\begin{equation}
  AD=\frac{1}{N}(\sum_{i=1}^{N}D_{i})
\end{equation}
where $D_{i}$ is the depth of session $i$ and $N$ is the number of sessions.

Moreover, in our online environment, we also evaluate the impact of the recommendation diversity, because the reason that our recommendation agent can keep users stay longer may be the increase of the recommendation variety which can discover the potential interests of users. Two metrics are adopted to evaluate the diversity, shown as follows.

\textbf{Category Coverage at Top-k Items(CC@K)}~\cite{wu2019pd}. It measures the average category coverage on top-k recommendation lists.
\begin{equation}\label{CC_k}
CC@k=\frac{1}{N}(\sum_{i=1}^{N}\frac{C_{k}}{\bm{C}})
\end{equation}
where ${C_{k}}$ is the number of categories covered by top-k items, $\bm{C}$ is the total number of categories available in the dataset and N is the number of sessions.

\textbf{Kullback-Leibler at Top-k Items(KL@K)}~\cite{steck2018calibrated}. It measures the distance of category coverage between clicked items and the items in current top-k list. It can be obtained as follows.
\begin{align*}
    p(c | u)=\frac{\sum_{i \in \mathcal{H}} w_{u, i}\hat{p}(c | i)}{\sum_{i \in \mathcal{H}} w_{u, i}},\quad q(c | u)=\frac{\sum_{i \in I} w_{r(i)} \hat{p}(c | i)}{\sum_{i \in I} w_{r(i)}}
\end{align*}
\begin{equation}
    \label{equ:q}    
    \tilde{q}(c | u)=(1-\alpha) \cdot q(c | u)+\alpha \cdot p(c | u)
\end{equation}
\begin{equation}
     C_{(p,q)}=C_{\mathrm{KL}}(p, q)=\mathrm{KL}(p \| \tilde{q})=\sum_{c} p(c | u) \log \frac{p(c | u)}{\tilde{q}(c | u)}
\end{equation}
\begin{equation}
     \mathrm{KL@K}=\frac{1}{N}(\sum_{i=1}^{N}C_{i}^{K}(p, q))
\end{equation}
where $\hat{p}(c | i)$ is the probability of category $c$ given item $i$, $p(c\/|u)$ and $q(c | u)$ are the overall probabilities of category $c$ in sets $\mathcal{H}$ and $I$ respectively, $H$ contains the clicked items before current recommendation, $I$ includes the items in current top-k list, $w_{u,i}$ and $w_{r(i)}$ are the weights of item $i$ (both of them are $1$ for simplification), $C_{\mathrm{KL}}(p, q)$ is the KL distance of category coverage between clicked items and the items in current top-k list, and $C_{i}^{K}(p, q)$ is the KL of i-th session in top-k list. Note that in case that $q$ is equal to $0$, we use $\tilde{q}$ instead of $q$, and $q \approx \tilde{q}$ with a small $\alpha>0$ as shown in Equation~\ref{equ:q}. Note that a lower KL-divergence value indicates the top-k list $I$ is more relevant to user's interest since user's interest can be revealed by historical clicked items $H$. .

\subsection{Experiment Settings}

\textbf{Evaluation on Yahoo}. For evaluation, we deal with Yahoo dataset as described in section 5.1, sequences, users' bounce positions and labels are generated naturally.
\textbf{Evaluation online}. If we use real world bounce labels in online scenarios , we can't evaluate when positions generated by agent exceed bounce labels. So we train a PBR model using the test set to simulate users' bounce rate at any position.

We use Adagrad optimizer to train simulation and agent models, where the embedding size $D$ is 700 in Yahoo and 128 in online scenarios. The $learning rate$ of the simulation model is set to $10^{-3}$ for Yahoo and $10^{-2}$ for online scenarios, respectively. The $learning rate$ of the agent model is set to $10^{-2}$ for Yahoo and $10^{-3}$ for online scenarios, respectively.

\subsection{Offline Experiments}


\subsubsection{Comparison with other models}

%

We compare our proposed model with the following representative methods~\footnote{In all experiments, we use a DeepFM\cite{guo2017deepfm} to deal with the user and items to get their embeddings.}.   

\textbf{Mart}. Mart\cite{friedman2001greedy} is a gradient boosting regression tree for ranking scenario.

\textbf{LambdaMart}. LambdaMart\cite{cao2007learning} is a list-wise rank method, it's the boosted tree version of LambdaRank, which is based on RankNet.

 \textbf{DNN}. DNN\cite{gupta2019architectural} is a widely used CTR model in real world E-commerce recommender systems, it can perform well without complex feature engineering.
    
\textbf{Context-Aware Transformer(CAT)}. CAT~\cite{pei2019personalized} is a multi-layer transformer model. To keep consistent, we do slight changes to address our input form, which equipped with personalized embedding to compute representations of initial input ranking list and output the re-rank scores. We list k items in descending order according to the score of auto-encoder framework in re-rank.
    
\textbf{Context-Aware LSTM based RNN Decoder (CAR)}. Inspired by \cite{GlobalRerank}, we uses RNN with LSTM to optimize CTR performance. CAR decodes items one by one, at $i$-$th$ step, it takes previous $i-1$ recommended items as the input and pickup the item with the highest predicted CTR from the RNN as the $i$-$th$ item, and so on. 

\textbf{Weighted Greedy Context-Aware RNN with CTR and PBR (WGCAR)}. It is extended from CAR. In addition to the original RNN responsible for CTR, an additional RNN is introduced for PBR. The recommended item is ranked according to the fusing score, $\alpha*CTR+(1-\alpha)*PBR$. The best  $\alpha=0.8$ for Yahoo and $\alpha=0.6$ for online dataset.
    
\textbf{Reinforcement Learning for CTE (RL4CTE)} The final version of our method, which treats CTE as the reward and CTE is obtained from the simulation environment. For each sequence, it uses the agent to sample N sequences. The sampled sequences are then used to calculate the baseline with the original sample and treated as the training samples.

\begin{table}[tb]

\scriptsize
\small
\caption{Comparison with other models}

\label{Tab03}

\begin{tabular}{ccccc}

\toprule

\multirow{2}{*}{method} & \multicolumn{2}{c}{Yahoo} & \multicolumn{2}{c}{Online}\\

\cmidrule(r){2-3} \cmidrule(r){4-5}

&AC    &AD

&AC      &AD \\

\midrule

$Mart$  &  0.89883  & 3.83976 & 0.89023 & 14.02381 \\

$LambdaMart$ & 0.88672  & 3.78733 & 0.88134 & 13.09132 \\

$DNN$          &0.92586            & 3.89558                             
& 0.93138         & 15.24645        \\

$CAT$         & 0.9362           &  3.98922                                
& 0.92106         & 14.94645       \\

$CAR$         &0.91405                & 3.84478
& 0.90225    & 14.14595          \\

$WGCAR$            & 0.93014                      & 4.09526                         
& 1.09875    & \textbf{18.54195}            \\

$RL4CTE$             & \textbf{0.97548}                       & \textbf{4.25875}                       & \textbf{1.18832}         & 18.4961           \\
    
\bottomrule

\end{tabular}

\end{table}


We compare our own method $RL4CTER$ with the baseline methods. The results are shown in Table~\ref{Tab03}. We can see that our proposed $RL4CTE$ performs much better than the baselines. In particular, our $RL4CTE$ defeats $WGCAR$, which indicates that it is better to optimize the unified goal CTE proportional to session clicks directly in RL than adopting the grid search of $\alpha$ to fuse CTR and PBR. 
Another observation is that WGCAR performs better than the other baselines. $WGCAR$ gets the best performance of AC at $\alpha=0.6$ instead of at $\alpha=1$, which shows that PBR is an important metric for session clicks.



\subsubsection{Effect of each component in our model}
We also conduct experiments on real-world recommendation system to investigate the effectiveness of each proposed components in our model, such as simulating environment, modified baseline and so on. Our method with different configurations are exhibited as follows. 


 \textbf{Static Data Learned REINFORCE (SDLR)}. In the absence of a simulation environment, we directly use the click and bounce labels provided by online dataset to calculate reward. For CTR, it equals 1 if the item has been clicked otherwise 0, and PBR is -1 if user bounce else 0.
 
\textbf{Simulation Environment Only with PBR (SE-PBR)}. The simulation environment only provides the estimated PBR, on the click side, the CTR signal is similar to SDLR, but the click label after bounce is forced to set as $0$. The final reward CTE is calculated with these two value.

\textbf{Simulation Environment Only with CTR (SE-CTR)}. on the contrary of SE-PBR, the simulation environment only provides the estimated CTR, and the bounce signal is set as $0.2$. The final reward CTE is calculated with these two value. 

\textbf{Simulation Environment wiht Whitening (SE-Whitening)}. In This method the simulation environment generate CTR and PBR for every item and get CTE reward via whitening.

\begin{table}[tb]

\scriptsize
\small
\caption{Ablation Study}

\label{Tab04}

\begin{tabular}{cccccccccc}

\toprule

\multirow{2}{*}{Method} & \multicolumn{2}{c}{Yahoo} & \multicolumn{2}{c}{Online}\\

\cmidrule(r){2-3} \cmidrule(r){4-5} 

&AC      & AD

&AC      & AD  \\

\midrule

$SDLR$            &  0.77802                    & 3.68454            
& 0.91763          & 17.24632                \\

$SE-PBR$           & 0.8436    & 3.89278                             & 0.92763          & 17.52487        \\

$SE-CTR$        &0.93236    & 3.98243                   & 0.9434     & 14.5069      \\
    
$SE-Whitening$     &  0.94831    & \textbf{4.34589}              &1.11452          & \textbf{18.55375}           \\

$RL4CTE$          & \textbf{0.97548}                       & 4.25875                    &\textbf{1.18832}          & 18.4961       \\

\bottomrule

\end{tabular}

\end{table}
%

 The experimental results are shown in Table~\ref{Tab04}. Firstly, the large margin between $RL4CTE$ and $SDLR$ demonstrates that the simulation environment can significantly improve the performance of our REINFORCE based method, since it can provide a reasonable interactive control. To be more specific, REINFORCE is an on-policy algorithm which requires a real-time interactive control to perform online update\cite{chen2019top}, but it is impossible to receive the true and real-time feedbacks from the static data. Leveraging the simulation environment, a virtual real-time interactive control can be achieved meeting the needs of REINFORCE.  


Besides, better performance of $RL4CTE$ than $SE-PBR$ indicates that the simulated CTR can efficiently deal with the short session and label missing problems. Concretely, in REINFORCE, given a fixed length trajectory, a shorter sequence is more likely to lose click labels since the click after bounce position is unobserved or unknown. Supplementing unknown values with predicted CTR labels can improve sample utilization. As a result, $RL4CTE$ can achieve better performance. The difference between $RL4CTE$ and $SE-CTR$ reveals that taking PBR into consideration is more efficient than taking cumulative clicks as reward. It also gives a further verification on PBR's importance for session clicks. In contrast with $RL4CTE$ and $SE-Whitenin$, the sampled baseline in $RL4CTE$ is better than the whitening baseline in $SE-Whitenin$.



\subsection{Online Experiments}
\textbf{Online deployment.} We can directly put the agent’s computation graph for online serving. The agent repeat $k$ times to generate final recommendation results. However, this process results in a very long response time(RT), which is not allowed for the online scenarios. There are two main factors for long RT. First, in each step, the encoder needs to recalculate the recommendation list from last step to get the correct state. Second, in the original graph, the fusion layer takes part in calculation which dramatically increase the time complexity. To tackle the issue, we take the advantage of the serial processing of RNN, and skip the recalculate process by restoring hidden state from the last timestamp. Further more, we split the fusion layer apart by storing and maintaining its output in a cache. By reorganizing the computation graph, the new graph is 14 times faster than the original one. In addition, we reduce the time complexity from $O(k^{2}n)$ to $O(kn)$, where $k$ is the recommended size and $n$ is the candidate size. The acceleration makes online serving possible.

\begin{table}[tb]
\small
  \caption{Online Results Comparison}
  \label{tab:freq}
  \begin{tabular}{cccc}
    \toprule
    \centering
    Method& $ \alpha $ & \multicolumn{1}{m{2.3cm}}{Relative Promotion of $AC$}
    & \multicolumn{1}{m{2.3cm}}{Relative Promotion of $AD$}\\
    \midrule
    \centering
    WG & 0.6 & \makecell[c]{$4.33\%$}& \makecell[c]{$3.10\%$}\\
    \centering
    SBR &- & \makecell[c]{$6.22\%$}& \makecell[c]{$4.92\%$}\\
    \bottomrule
  \end{tabular}
\end{table} 

\textbf{Online A/B tests}. Table 4 shows that $RL4CTE$ obtains a relative promotion of $6.22\%$ at AC and $4.92\%$ at AD than the baseline without re-ranking. In addition, it performs better than the best performance of $WGCAR$. This proves $RL4CTE$ also works well for real-world E-commerce scenario. 

In addition, as described in the dataset section, diversity is an important metric for average clicks. Therefore, we analysis the performance of the models in CC and KL mentioned before.
\begin{table}[tb]
\small
\caption{PBR with CC and KL}
\label{tab:diversity}
\begin{tabular}{cccccc}
    \toprule
    Method                                                     & $\alpha$     & CC@5            & CC@10           & KL@5             & KL@10           \\
    \midrule
    \multicolumn{1}{c}{$RL4CTE$} & -            & \textbf{0.03788}         & \textbf{0.06476}          & \textbf{3.12958}          & \textbf{2.88818}         \\
    \midrule
    \multirow{6}{*}{$WGCAR$}                          & 0.0          & 0.038405          & 0.06397          & 2.94835          & 2.77452         \\
    & 0.2          & 0.03424          & 0.05757         & 3.05903          & 2.79424         \\
    & 0.4          & 0.03121         & 0.05536          & 3.12354          & 2.84444         \\
    & \textbf{0.6} & \textbf{0.02948} & \textbf{0.05334} & \textbf{3.15241} & \textbf{2.8917} \\
    & 0.8          & 0.02870          & 0.05201          & 3.17702          & 2.92983         \\
    & 1.0          & 0.02818          & 0.05088          & 3.18800             & 2.95879  \\
    \bottomrule
\end{tabular}
\end{table} \\

The results are shown in Table~\ref{tab:diversity}. We can see that the personal diversity is proportional to PBR as lower $\alpha$ indicates larger PBR shown in the results of $WGCAR$. KL decreases as CC increase, which reveals that the explicit consideration of PBR is able to enhance personal diversity and still matches user interests. Compared with the best performance of $WGCAR$ when 
$\alpha=0.6$, $RL4CTE$ improves the AC by $2.89\%$. In addition, the performance of $RL4CTE$ in terms of CC and KL is also better. This demonstrates that $RL4CTE$ is able to explore the diversity of items and matching the interests of users effectively.

\section{Conclusion}
\label{sec:con}

In this paper, we have proposed a reinforcement learning based re-ranking model to optimize average number of clicks in a session. In our re-ranking method, we used transformer framework to build a simulation environment and trained a RNN based re-ranking agent using REINFORCE method. We have demonstrated that our re-ranking method can boost performance on both offline public dataset and real-world online scenario.

\bibliographystyle{ACM-Reference-Format}
\bibliography{sample-base}

\appendix

\end{document}